\documentclass[ag]{copernicus}

\begin{document}

\title{Collisionless reconnection: Mechanism of self-ignition in thin current sheets 
}

\author[1]{R. A. Treumann\thanks{Visiting the International Space Science Institute, Bern, Switzerland}}
\author[3]{R. Nakamura}
\author[3]{W. Baumjohann}

\affil[1]{Department of Geophysics and Environmental Sciences, Munich University, Munich, Germany}
\affil[]{Department of Physics and Astronomy, Dartmouth College, Hanover NH 03755, USA}
\affil[3]{Space Research Institute, Austrian Academy of Sciences, Graz, Austria}

\runningtitle{Seeding reconnection}

\runningauthor{R.A. Treumann, R. Nakamura, and W. Baumjohann}

\correspondence{R.A.Treumann\\ (rudolf.treumann@geophysik.uni-muenchen.de)}

\received{ }
\revised{ }
\accepted{ }
\published{ }


\firstpage{1}

\maketitle

\begin{abstract}
The spontaneous  onset of magnetic reconnection in thin collisionless current sheets is shown to result from a thermal-anisotropy driven magnetic Weibel-mode, generating seed-magnetic field {\sf X}-points in the centre of the current layer.

\end{abstract}

\introduction
The idea of magnetic reconnection as the main plasma process that converts stored magnetic energy into kinetic energy originates from the intuitive geometric picture of annihilating antiparallel magnetic field lines when approaching each other  \citep[see, e.g.,][]{sweet1957,parker1958,dungey1961}.  Observations in space  have unambiguously confirmed the presence of reconnection under collisionless conditions \citep[see, e.g.,][]{fujimoto1997,oieroset2001,nagai2001} when the fluid theoretical approaches break down. However, no convincing theoretical argument for the spontaneous occurrence of reconnection has so far been given. In collisionless numerical simulations reconnection is artificially ignited \citep[cf., e.g.,][]{zeiler2000}, mostly by {\it ad hoc} imposing a seed {\sf X}-point in the current sheet separating the anti-parallel fields. The ongoing search for the mechanism of spontaneous onset of collisionless reconnection points to the `missing microphysics' in thin current sheets. 

In the present Letter we show that instability of the inner current layer gives rise to the self-consistent generation of local magnetic fields ${\bf B}=(B_x,0,B_z)$ transverse to the current layer.  Such local fields are equivalent to the generation of microscopic seed {\sf X}-points in the current sheet centre and are capable of spontaneously igniting reconnection as is known from two-dimensional PIC particle simulations. 
Since in an ideal current sheet ions and electrons become non-magnetic on their respective inertial scales $\lambda_{\,i,e}=c/\omega_{\,i,e}$, where $\omega_{\,i,e}=e\sqrt{N/\epsilon_0m_{\,i,e}}$ are the plasma frequencies of ions and electrons, (classical) collisionless convective transport of magnetic fields into the current layer takes place up to a vertical distance $z\sim\lambda_{\,e}$ from the centre of the current sheet. The region between $\lambda_{\,e}\lesssim z\lesssim\lambda_{\,i}$ is known as the `Hall-current' \citep{sonnerup1979} or (mistakenly, as there is no diffusion present)  `ion-diffusion' region. Being a by-product of thinning of the current layer, the Hall currents are -- presumably -- not involved in the reconnection process proper.\footnote{The question of the role of Hall currents in reconnection is not resolved yet. Classically they are unimportant for the reconnection process. It is, however, not certain whether or not on the microscopic scales non-classical (quantum-Hall) effects are induced by the environmental conditions \citep[einselection effects, see][]{zurek2003} in which Hall-electrons would more directly be involved.} They close along the magnetic field by electrons that are accelerated in the oblique lower-hybrid-drift/modified-two-stream instability driven by magnetised Hall-electrons on the non-magnetic ion background thereby coupling the reconnection site to the auroral ionosphere \citep{treumann2009}. 

When speaking of a current sheet, we refer to ideal current sheets separating strictly  antiparallel fields. The observational paradigm of a reconnecting current sheet is the magnetospheric tail-current sheet. This current sheet is not ideal in the above sense as it is embedded into a quasi-dipolar field which still might preserve a weak rudimentary (normal) magnetic field component $B_z$pointing northward. This $B_z$ component re-magnetises the central-sheet electrons and affects the evolution of (collisionless) tearing modes \citep{galeev1975}. Nevertheless below, when using numbers, we will for reasons of resolution refer to conditions in the magnetotail even though our theory might better apply to the magnetopause, interplanetary space or astrophysics.  In principle, observation of the electron-inertial (`electron-diffusion') region is difficult because of its narrow width. Unambiguous observations do not yet exist. At the magnetopause, in particular, very narrow electron layers have sometimes been reported assuming that they relate to the electron-inertial region during reconnection \citep[for a recent discussion of the experimental prospects of resolving the electron-inertial region cf.,][]{scudder2008}.

\section{Magnetic field generation in the current layer}
Unless a guide field is imposed from the outside, the inner current region $z\lesssim\, {\rm few}\,\lambda_e$ is about free of magnetic fields, 
while at the same time carries a (diamagnetic) current ${\bf J}_\perp$ perpendicular to the antiparallel magnetic fields to both sides of the current, caused (for instance in the geomagnetic tail current sheet or the Earths magnetopause) by a (macroscopic) electric  potential drop $\Delta \,U$ along the current. 

For the understanding of the mechanism of  reconnection it is of no interest how this potential drop is generated. This may happen when two magnetised collisionless plasmas of finite lateral extension collide. In the magnetotail current sheet the potential amounts to $1\lesssim\Delta\, U\lesssim$ few 10\,kV, and electron and ion temperatures are of the order of $T_e\sim 0.1$ keV and $T_i\sim 1$ keV, respectively. 
Electrons entering the centre of the current sheet accelerate along the current, thereby becoming the main current carriers here. Their high translational velocity $V_e=\sqrt{e\Delta\, U/m_e}>v_e$ 
exceeds their thermal speed $v_e=\sqrt{2T_e/m_e}$ providing conditions that are unstable against the Buneman two-stream instability \citep{buneman1958}, a fast growing electrostatic instability with high frequency  $\omega_{\,B}\sim 0.03\omega_e$ and large growth rate $\gamma_{\,B}\sim\omega_{\,B}$ \citep[cf., e.g.,][p. 22]{treumann1997}. In the geomagnetic tail current sheet the growth rate amounts to $\gamma_{\,B}\approx 1.7$ kHz, corresponding to a growth time of $\tau_{\,B}\sim$ 0.006 s. 

The Buneman instability readily generates localised electrostatic structures (known as electron and ion phase space holes) which trap a substantial part of the electrons and heat them in the direction along the current drift velocity. Numerical simulations suggest that this process takes roughly 100-1000 plasma periods \citep{buneman1959,newman2002}, or few 10 e-folding times, in the magnetospheric tail $\lesssim 0.1$ s. In this process the instability shuts off itself by increasing the parallel electron temperature until $v_{e\|}\sim V_e$. At the end of this very fast process the electrons develop a  temperature anisotropy 
\begin{equation}
A=T_{e\|}/T_{e\perp}-1>0
\end{equation}
with current-parallel temperature $T_{e\|}>T_{e\perp}=T_e$ exceeding the initial electron temperature, roughly $A\lesssim 1$ in the magnetospheric tail current sheet. The subscripts $|\!|$ and $\perp$ refer to the respective directions of maximum and minimum electron temperatures, i.e. the two directions of the electron pressure tensor
\begin{equation}
{\sf P}_e=N[T_{e\perp}{\sf I}+(T_{e\|}-T_{e\perp}){\bf V}_e{\bf V}_e/V_e^2]
\end{equation}

In this thermally anisotropic case the electrons obey a bi-Maxwellian equilibrium distribution function
\begin{equation}\label{eq1}
f_e(v_\perp,v_\|)=\frac{(m_e/2\pi )^{\!\!\frac{3}{2}}}{T_{e\perp}\sqrt{T_{e\|}}}\exp\left[-\frac{m_ev_\perp^2}{2T_{e\perp}}-\frac{m_ev_\|^2}{2T_{e\|}}\right]
\end{equation}
which, in a nonmagnetised plasma (like the inner current region $z\lesssim\lambda_e$) is unstable with respect to the family of Weibel\footnote{Weibel -- or current filamentation instabilities, as they are sometimes called following \citet{fried1959} where a simple physical model of their mechanism was given early -- have mostly been investigated in view of astrophysical applications in a relativistic approach.} instabilities \citep{weibel1959}. These are very low (about zero) frequency (purely growing) electromagnetic instabilities which are capable of generating {\it stationary} magnetic fields that grow from thermal fluctuations (not requiring any magnetic dynamo mechanism).
The linear electromagnetic dispersion relation of the plasma  becomes
\begin{equation}\label{eq2}
(n^2-\epsilon_{\perp})^2\epsilon_{ \ell}=0
\end{equation}
where $n=kc/\omega$ is the refraction index, and $\omega$ is the frequency of the linear disturbance. The dielectric tensor has the two scalar components $\epsilon_{ \ell}({\bf k},\omega), \epsilon_\perp({\bf k},\omega)$ which are the longitudinal and transverse response functions, respectively. For our purposes it suffices to consider the electromagnetic (transverse) response buried in 
\begin{equation}\label{eq3}
\epsilon_\perp= 1-\frac{\omega_e^2}{\omega^2} \left\{ 1-(A+1) \left[ 1+\zeta Z(\zeta) \right] \right\} -\frac{\omega_i^2}{\omega^2}=n^2
\end{equation}
where $Z(\zeta)$ is the plasma dispersion function, $\zeta=\omega/k_\perp v_{e\perp}$, and $v_{e\perp}=\sqrt{2T_\perp/m_e}$ is the electron thermal speed perpendicular to the current. The Weibel instability grows in the plane perpendicular to the direction of higher thermal velocity, which in our case has been assumed as the parallel direction. Hence, ${\bf k}=(k_x,0,k_z)=(k_\perp\sin \theta, 0, k_\perp\cos\theta)$; in an extended medium there is no $\theta$-dependence, a point to which we will return later.  The contribution of the resting ions has been retained for completeness; because of the smallness of the ion plasma frequency $\omega_{\,i}\ll\omega_{\,e}$, being much less than the electron plasma frequency $\omega_{\,e}$, it plays no role in the instability. 
\begin{figure}[t!]
\centerline{{\includegraphics[width=0.4\textwidth,clip=]{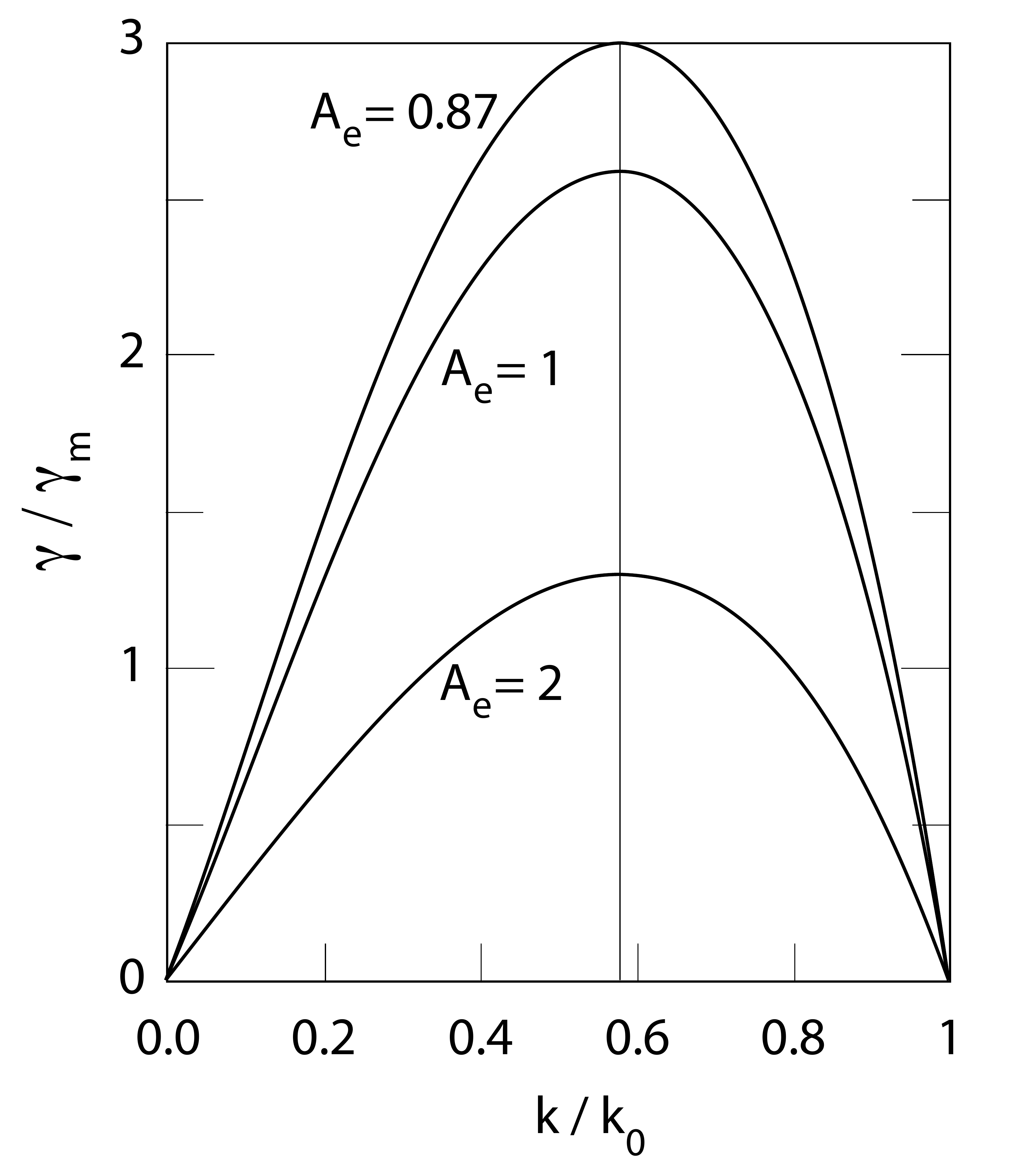} }}
\caption[ ]
{\footnotesize The anisotropic-thermal Weibel instability growth rate $\gamma/\gamma_m$, normalised to maximum growth, as function of the normalised wavenumber $k/k_0$ for three different thermal anisotropies. This ratio increases as $A^{-1}$. The vertical line indicates the position of the maximum growing wave number $k_m/k_0$.}\label{weibel-2}
\vspace{-0.3cm}
\end{figure}

At zero real frequency $\omega=i\gamma$ and $A>0$ the right-hand side of Eq.\,(\ref{eq3}) becomes the dispersion relation of the thermal-anisotropy driven Weibel mode \citep[][and others]{weibel1959,yoon1987}.  Instability $\gamma(k_\perp)>0$ sets on at phase velocities $\omega/k_\perp\ll v_{e\perp}$ for wavenumbers $k_\perp<k_{\rm 0}$, 
\begin{equation}
k_{0}\lambda_e\simeq \sqrt{A}
\end{equation}
with instability growth rate
\begin{equation}
\frac{\gamma_{\,\rm W}}{\omega_e}\simeq \sqrt{\frac{2}{\pi}}\frac{v_{e\perp}}{c}\frac{k_\perp}{k_0}\left( 1- \frac{k_\perp^2}{k_{\rm 0}^2}\right)(A+1)(k_{\rm 0}\lambda_e)^3 
\end{equation}
vanishing at long wavelengths $k_\perp=0$. The growth rate maximises at wavenumber $k_{\perp{\rm m}}=k_0/\sqrt{3}=\lambda_e^{-1}\sqrt{A/3}$ (see Figure \ref{weibel-2}) where its value is 
\begin{equation}
\frac{\gamma_{\,\rm W,m}}{\omega_e}\simeq \frac{4}{3}\sqrt{\frac{A^3\Theta_{e}}{3\,\pi}}(A+1)
\end{equation}
with $\Theta_e\equiv T_{e\perp}/m_ec^2$ the (ambient) temperature normalised to the rest energy of an electron. Numerically this expression yields for the maximum growth rate
\begin{equation}
\gamma_{\,\rm W,m}\approx 34 \sqrt{N_{\rm [cm^{-3}]}T_{e\perp[{\rm eV}]}}A^\frac{3}{2}(A+1) \ \ {\rm Hz}
\end{equation}
Depending on the value of the anisotropy, this growth rate can be substantial.
If $A>1$, it grows as $\gamma\propto \sqrt{A^5}$, while for anisotropies $A<1$ it grows like $\gamma\propto \sqrt{A^3}$.  In the tail plasma sheet we have $T_e\sim 100$ eV and $N\sim 1$ cm$^{-3}$. Then, even with $A\sim 0.1$ one finds quite a fast growth rate of $\gamma_{\,\rm W,m}\lesssim 10$ Hz.

The important point is that even though the growth rate might not be extraordinarily large, it generates a magnetic field that has two components, ${\bf B}_{\rm W}=(B_x, 0,B_z)$, both being transverse to the initial current. The component $B_x$ is alternating between the directions parallel and antiparallel to the initial magnetic field outside the current layer, being directed $\pm\hat x$ while the other component is perpendicular to the current layer directed along $\pm \hat z$. This field modulates the current layer along $\hat x$ causing magnetic islands whose vertexes lie in the centre of the current layer. It thus provides seed-{\sf X} points which, if sufficiently large amplitude, will spontaneously ignite reconnection. The finite nonmagnetic current sheet width in $z$ imposes a limit $2\pi/k_z < 2\lambda_e$ which yields
\begin{equation}
k_z/k_x={\rm cot}\,\theta\approx k_z/k_m>\pi\sqrt{3/A}
\end{equation}
the lower limit resulting from the restriction on $A>3m_e/2m_i$ (see below). Thus the Weibel mode propagates at angles 
\begin{equation}
{\rm tan}^{-1}[\pi^{-1}\sqrt{m_e/2m_i}] <\theta<{\rm tan}^{-1}[\pi^{-1}\sqrt{A/3} ] 
\end{equation}
against ${\hat x}$.  This is the maximum angle the wavevector assumes in the Weibel-field vertexes. For $A=0.1$ and $A=1$ this inclination angles are $0.3^\circ<\theta\lesssim 3.4^\circ$ and $\sim11^\circ$, respectively. However, in addition, the Weibel mode can propagate in two directions $\pm{\hat x}$. The two cases are shown in Figure \ref{weibel-1}:  ($a$) when the propagation direction choses to be along the external field. In this case simple seed-{\sf X} points in the current sheet are generated which will allow reconnection to evolve in the usual way. For the oppositely directed Weibel vertices shown in Figure \ref{weibel-1} ($b$), however, a multitude of additional reconnection sites are produced along $z\sim\pm\lambda_e$, and the current layer becomes highly unstable.
Which is the most probable case can be decided only after a complete solution of the Weibel-unstable boundary value problem of the current layer.
\begin{figure}[t!]
\centerline{{\includegraphics[width=0.5\textwidth,clip=]{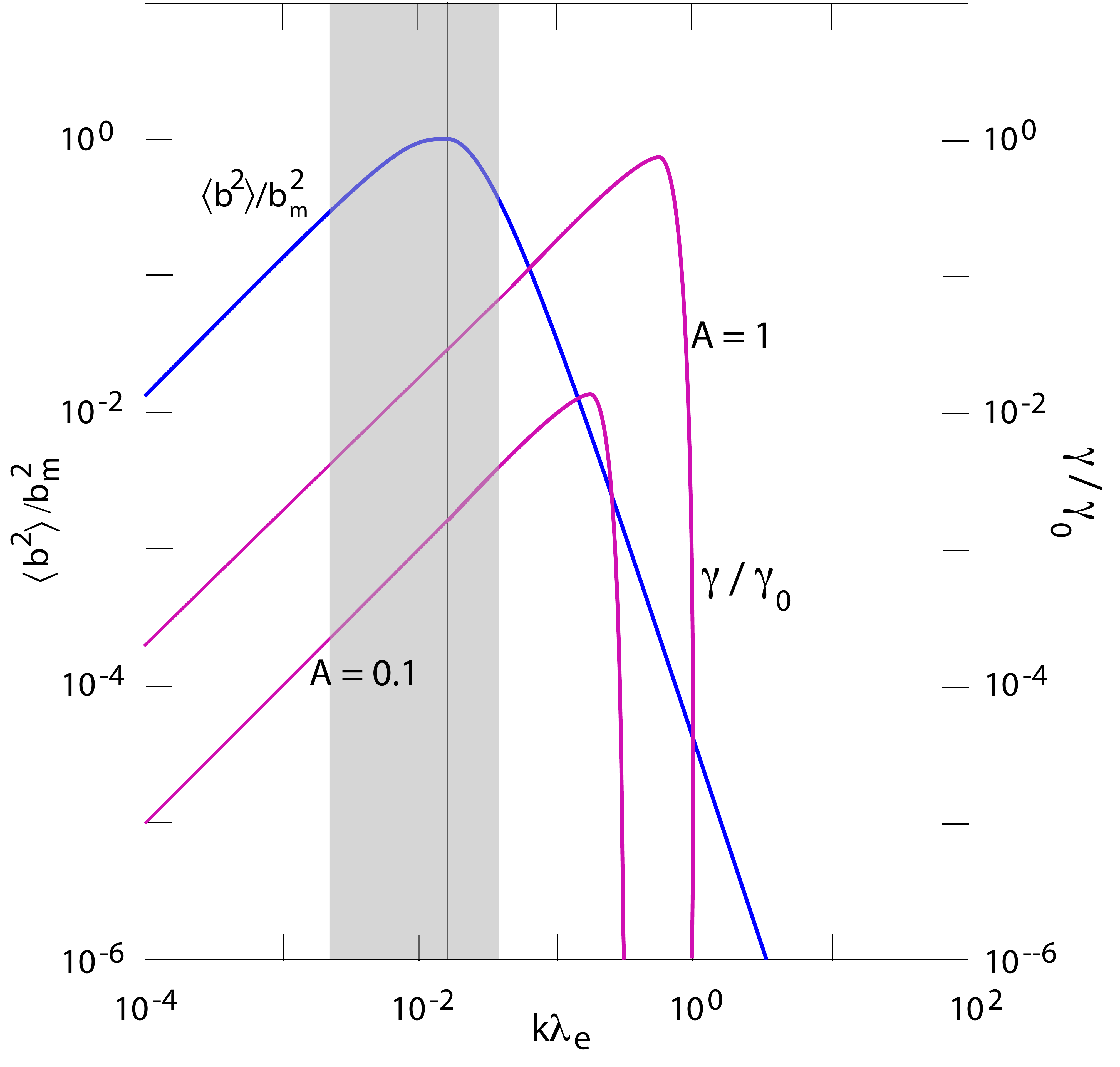} }}
\caption[ ]
{\footnotesize  Wavenumber dependencies of the normalised fluctuation spectrum and normalised growth rates. The normalisation of the thermal fluctuation spectrum is to its maximum value given in Eq. (\ref{eq13}). Normalisation of the growth rate is to $\gamma_0=\omega_e\sqrt{\pi/2}(c/v_e)$. Since the growth rate depends on anisotropy $A$ it is given for the two cases $A=0.1, 1$. Note the competition between growth rate and fluctuation level. At long wavelengths the high fluctuation level partially compensates for the low growth rate. The range of wavelength of interest in the magnetospheric tail is shown shaded. It centres around maximum thermal fluctuation level.}\label{weibel-6}
\vspace{-0.3cm}\label{weibel-fig6}
\end{figure}

\section{Thermal fluctuation level}
In order to infer how long it takes the instability to achieve substantial magnetic field amplitudes we need to estimate the magnetic thermal fluctuation level $\langle b_ib_j \rangle_{k,\omega=0}$ from where the Weibel instability starts growing in the presence of the electron pressure anisotropy (thermally fluctuating magnetic fields will be denoted by lower case letters). Magnetic thermal levels have recently been estimated \citep{yoon2007,baumjohann2010}. From basic fluctuation theory \citep{sitenko1967} the spectral energy density of the zero-frequency thermally-anisotropic Weibel mode can be written
\begin{equation}\label{eq12}
\frac{\langle \left| {\bf b}\right|^2\rangle_{k0}}{\sqrt{2\pi}} =\frac{\mu_0}{\omega_e}\frac{c}{v_{e\perp}}
\frac{T_{e\perp}k_\perp\lambda_e(A+1)^2}{(A+2)[k_\perp^2\lambda_e^2-A+m_e/m_i]^2}
\end{equation}
The 0-subscript refers to vanishing real frequency. Here the ion contribution has been retained. In the isotropic $A=0$ and Weibel-stable $-2<A<0$ cases, the spectral energy density vanishes at $k_\perp\to 0, k_\perp\to\infty$ and, in a proton-electron plasma, maximises at $k_\perp\lambda_e\approx0.013$. Its maximum value is
\begin{equation}\label{eq13}
\langle \left| {\bf b}\right|^2\rangle_{k0,m} =8.25\times10^{-23}\,\sqrt{\frac{T_{e[{\rm eV}]}}{N_{[\rm cm^{-3}]}}}\quad {\rm\frac{V^2s^3}{m}}
\end{equation}
\begin{figure}[t!]
\centerline{{\includegraphics[width=0.45\textwidth,clip=]{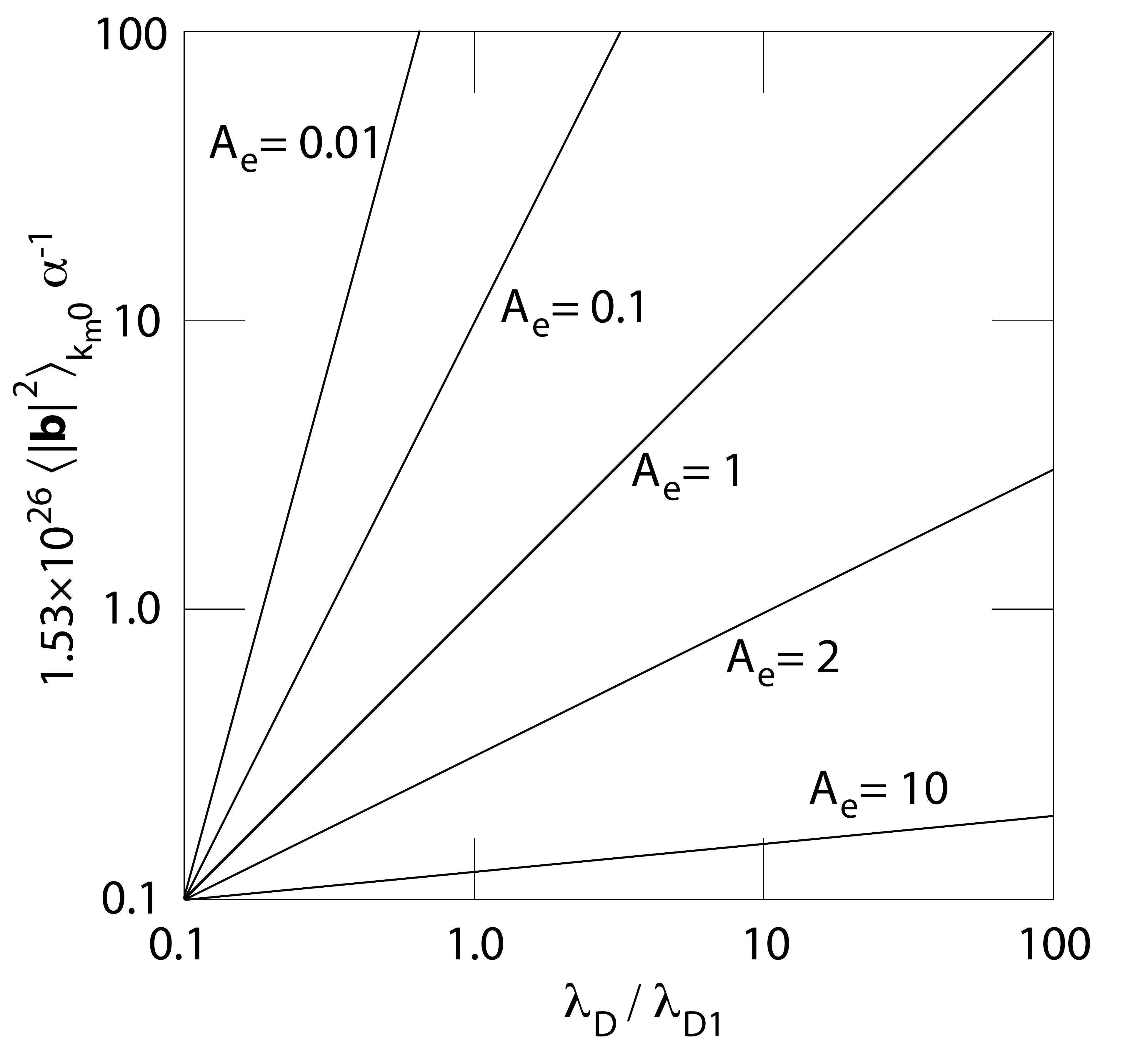} }}
\caption[ ]
{\footnotesize Normalised thermal fluctuation level at maximum growing Weibel wave number $k_m$ as function of Debye length $\lambda_D$. The fluctuations are normalized to their value at $T_e=$
 1 eV, $N$ = 1 cm$^{-3}$. The Debye normalisation is taken to the Debye length $\lambda_{D1}$ at these numbers. As suggested by Fig. \ref{weibel-fig6}, the initial fluctuation level from where the maximum unstable Weibel mode grows decreases  $\sim A^{-1}$ because of its dependence on $k_m$, which increases as $\sqrt{A}$. Since the spectral energy density of fluctuations decreases  $\sim k^{-3}$, the initial level of the fastest growing Weibel mode also decreases with growing anisotropy. }\label{weibel-3}
\vspace{-0.3cm}
\end{figure}

One might note that for positive anisotropies the current sheet is not in equilibrium anymore, and the thermal fluctuations explode close to the boundary of the unstable domain for $k_\perp\lambda_e\sim \sqrt{A}\approx k_0\lambda_e$ indicating onset of instability and phase transition.

\subsection{Fastest growing Weibel mode}
The Weibel instability choses from this spectral energy density and supports the fastest growing wavenumber $k_{\perp{\rm m}}$. Inserting for $k_{\perp{\rm m}}$ the initial thermal level of the fastest growing mode becomes
\begin{equation}
\langle|{\bf b}|^2\rangle_{k_{\rm m}0}\simeq \frac{9\mu_0}{4}\frac{m_ec^2}{\omega_e}\sqrt{\frac{\pi}{3}\frac{T_{e\perp}}{m_ec^2}}\frac{(A+1)^2/(A+2)}{(A-3m_e/2m_i)^2}
\end{equation}
Since large thermal anisotropies are unrealistic, the cases of small  $A\ll 1$ and large anisotropies $A\sim 1$ may be distinguished yielding the limiting initial levels
\begin{equation}
\langle |{\bf b}|^2\rangle_{k_{\rm m}0}\simeq \frac{\alpha\mu_0}{A}\frac{m_ec^2}{\omega_e}\sqrt{\frac{\pi}{3}\frac{T_{e\perp}}{m_ec^2}}, \quad A>\frac{3}{2}\frac{m_e}{m_i}
\end{equation}
with $\alpha=9/8$ for $A\ll 1$, and $\alpha=3$ for $A\lesssim 1$. Numerically:
\begin{equation}
\langle |{\bf b}|^2\rangle_{k_{\rm m}0}\approx 8.8\times 10^{-28}\frac{\alpha}{A}\sqrt{\frac{T_{e[{\rm eV}]}}{N_{\rm [cm^{-3}]}}} \quad \frac{\rm V^{\,2}\,s^3}{\rm m}
\end{equation}
where the temperature is measured in eV, and the density is in cm$^{-3}$. The numerical factor for the largest expected anisotropy $A\sim 1, \alpha=3$ is $\approx 2.63\times10^{-27}$. 

\begin{figure}[t!]
\centerline{{\includegraphics[width=0.5\textwidth,clip=]{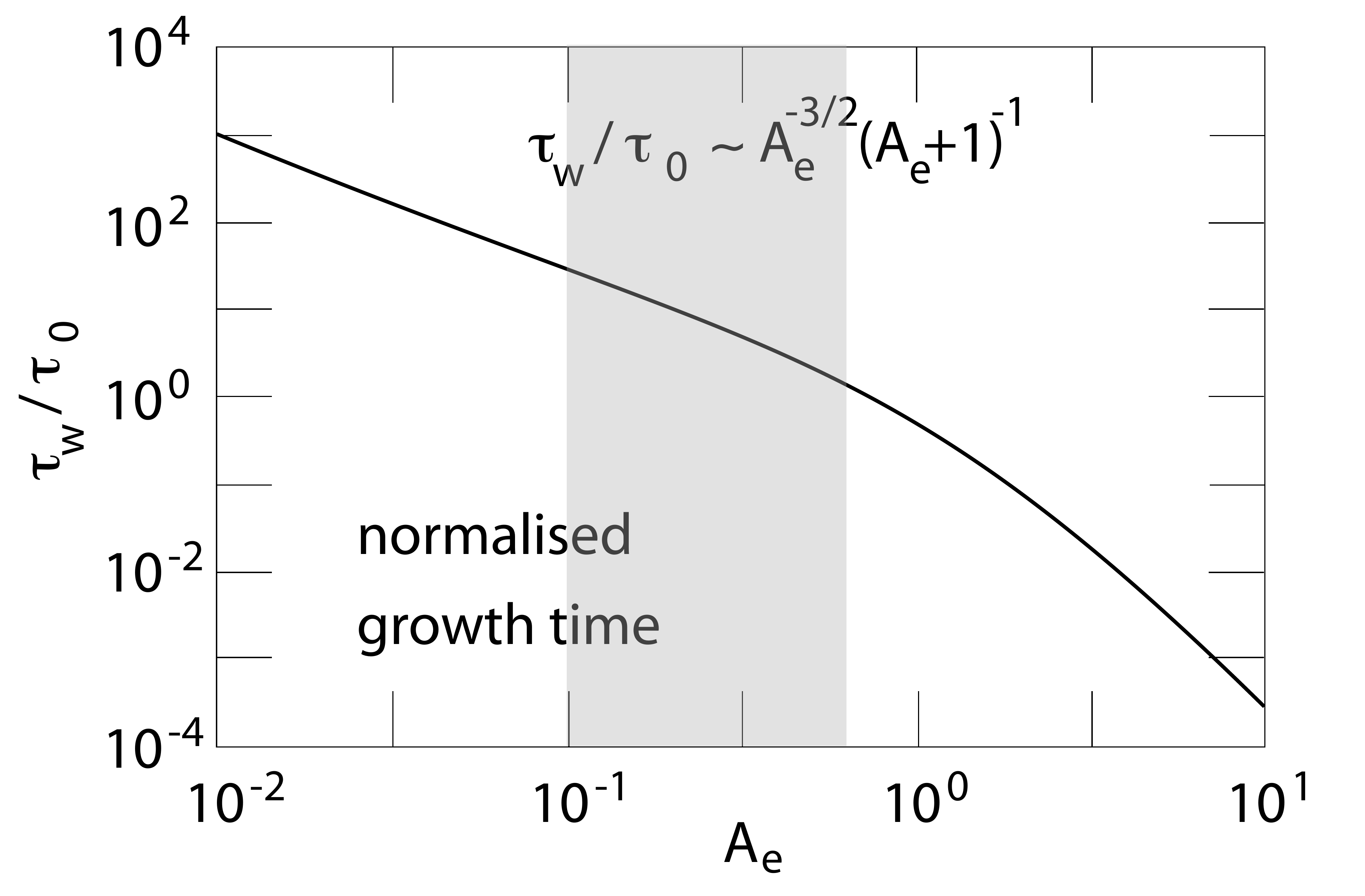} }}
\caption[ ]
{\footnotesize Variation of the relevant growth times $\tau_{\, \rm W,m}$ (Eq. \ref{eq-13}) of the maximum growing Weibel mode with thermal anisotropy $A$ for maximum growth. The normalisation $\tau_0$ is a number which depends on $T_e, N_e$ and the value of the final  magnetic field strength required  for onset of reconnection. In the geomagnetic tail current sheet $\tau_0\gtrsim 0.1$ s. Even though the initial thermal level in Fig. 2 decreases with anisotropy, the times to reach a substantial Weibel field strength in the current sheet are mainly determined by the dependence of the maximum growth rate $\gamma_m$ on $A$. They depend  only logarithmically on the initial thermal wave field. They thus decrease with growing anisotropy. The shaded area shows the range of anisotropies and growth times expected in the geomagnetic tail current sheet. }\label{weibel-4}
\vspace{-0.3cm}
\end{figure}
The unstable Weibel spectral energy density evolves according to
\begin{equation}
\langle |{\bf B}(t,  k_{\rm m},0)|^2\rangle\approx \left\langle|{\bf b}|^2\right\rangle_{k_{\rm m}0}\exp (2\gamma_{\,\rm W,m} t)
\end{equation}
The growth time of the fastest growing mode follows from this expression as
\begin{equation}\label{eq-13}
\tau_{\,\rm W,m}\approx \frac{1}{2\gamma_{\,\rm W,m}}\ln\frac{\langle|{\bf B}(k_{\rm m},\tau_{\,\rm W,m})|^2\rangle}{\langle|{\bf b}|^2\rangle_{k_{\rm m 0} }}
\end{equation}
The spectral energy density of a $|{\bf B}|=$ 1 nT magnetic field fluctuation is $\langle |{\bf B}_{\rm 1 nT}|^2 \rangle_{k0} \approx 4.3\times 10^{-12}\,\, {\rm V^2\,s^3 /m}$. This value may be used when estimating the time it needs for the maximum growing thermal-anisotropy driven Weibel mode in the magnetotail current sheet to grow up to a value comparable to the external (lobe) magnetic field $B_0\sim$ few nT. If we take the growth rate in the range $1\lesssim\gamma_{\,\rm W,m}< 50$ Hz which holds for $0.1\lesssim A<1$, short growth times from thermal level to 1 nT fields of the order of 
\begin{equation}
\tau_{\,\rm W,m}>0.1\quad{\rm s}
\end{equation}
are obtained, corresponding to mostly a few seconds of growth time in the magnetospheric tail. Given the uncertainty of the numerical values used, this is not an unreasonable estimate of the length of the ignition phase that initiates reconnection in the tail current sheet, i.e. the time to produce initial {\sf X}-points which subsequently start reconnection. Typical times for the evolution of substorms following onset of reconnection range from minutes to few tens of minutes and depend on the connection of the magnetotail reconnection site to the response of the ionosphere. 
\begin{figure}[t!]
\centerline{{\includegraphics[width=0.4\textwidth,clip=]{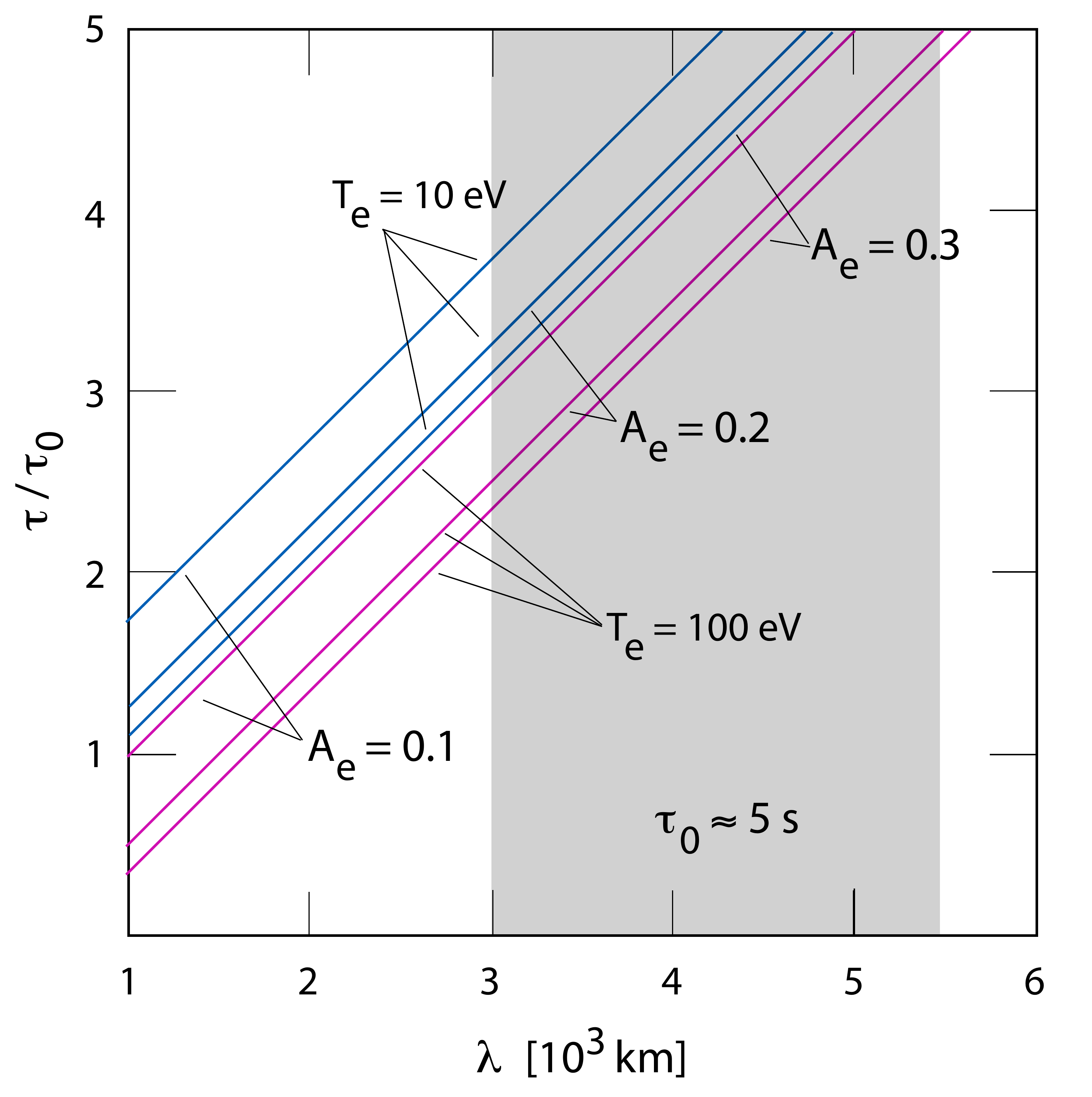} }}
\caption[ ]
{\footnotesize The Weibel mode at long wavelengths: Growth time dependence on wavelength in the long wavelength range. The normalisation $\tau_0\approx 5$ s is obtained for $T_e=100$ eV, $N_e= 1$ cm$^{-3}$, and $A=0.1$. Growth times are shown for two different temperatures $T_e=10$ eV and $T_e=100$ eV and three values of the anisotropy $A=0.1, 0.2, 0.3$. The shaded area is the acceptable wavelength domain for the geomagnetic tail.}\label{weibel-fig5}
\vspace{-0.3cm}
\end{figure}
\subsection{Long wavelength Weibel modes}
Fastest growth corresponds to very short wavelengths $k_\perp\lambda_e\lesssim\sqrt{A/3}$. There may, however, as well be competition between decreasing growth rate and increasing initial fluctuation level at long wavelengths as shown in Fig. \ref{weibel-fig6}.

Equation\,(\ref{eq12}) suggests that the spectral energy density of thermal fluctuations for $k^2_\perp\lambda_e^2<A+m_e/m_i$ increases as  $\sim k_\perp\lambda_e$. In isotropic plas\-ma $A=0$ this implies wavelengths $\lambda\gg 2\pi\lambda_e\sqrt{m_i/m_e}\approx 300\lambda_e$. In the magnetospheric tail the wavelength of maximum thermal fluctuation level is thus $\lambda\sim 1500$ km. The spectral energy density in this long-wavelength range is given by Eq. (\ref{eq13}).

At such wavelengths one can neglect the term $k_\perp/k_0$ in the expression for the Weibel growth rate. For small aniso\-tropies $A<1$ the growth rate becomes
\begin{equation}
\gamma_W\simeq \frac{4A c}{\lambda}\sqrt{\frac{\pi T_{e\perp}}{m_ec^2}}
\end{equation}
Inserting the long-wavelength restriction on $\lambda$ yields
\begin{equation}
{\gamma_W}\!\ll\!\!0.01A\sqrt{\frac{\pi T_{e\perp}}{ m_ec^2}}{\omega_e}\approx 1.4A\sqrt{T_{e[\rm e\!V]}N_{[\rm cm^{\!-3}]}}~~{\rm Hz}
\end{equation}
The Weibel growth rate, for $A\sim 0.1$ and $T_e\sim 0.1$ ke$\!$V, in this wavelength range is thus of the order of $\gamma_W\sim 0.1$ Hz, one order of magnitude less than at maximum growth, yielding exponentiation times $\gamma_W^{-1}\sim 1$ s and growth times $\tau_W\sim 10$ s (see Fig. \ref{weibel-fig5}). This is the time a Weibel wavelength of $\lambda\sim 1500-3000$ km, i.e. roughly half one Earth radius, needs to grow from thermal level to an amplitude of 1 nT in the geomagnetic tail prior to onset of reconnection. 

In anisotropic plasma $A\neq 0$ and we may relax the condition on the wavelength. In this case the mass ratio in the thermal fluctuation expression becomes unimportant for reasonably large $A\gg m_e/m_i$. Then long wavelengths imply that $k_\perp\lambda_e\ll\sqrt{A}=k_0\lambda_e$ and
\begin{eqnarray}
\langle \left| {\bf b}\right|^2\rangle_{k0}&\simeq&\frac{\mu_0m_ec^2}{\omega_e}\sqrt{\frac{\pi T_{e\perp}}{m_ec^2}}\frac{k_\perp\lambda_e}{A^2}\nonumber \\
&\ll& \frac{\mu_0m_ec^2}{\omega_e}\sqrt{\frac{\pi}{A^3}\frac{T_{e\perp}}{m_ec^2}}\\
&\approx&3.2\times10^{-24}{A^{-\frac{3}{2}}}\sqrt{{T_{e[{\rm e\!V}]}}}\quad{\rm{V^2s^3}/{m}}\nonumber
\end{eqnarray}
With $T_{e\perp}=100$ eV, and $A=0.1$ and using the former expression for the growth rate, one correspondingly expects growth times  from thermal level of the order of $\tau_W\sim 100$ s, between 1 and 2 min, for wavelengths of the order of $\lambda\sim$\,1000\,km $\gg 2\pi\lambda_e/\sqrt{A}\sim 110$\,km.

\section{Collisionless reconnection scenario} 
These estimates are sufficiently encouraging for developing a microscopic scenario for collisionless reconnection as follows: 
Assume a plane Harris current layer $J_y=-J_0{\rm sech}^2(2z/\Delta)$, with $\Delta$ the layer half-width, separating two (lobe) regions of antiparallel magnetic fields. The magnitude of the field changes as $B_x(z)=B_0{\rm tanh}(2z/\Delta)$. Let this current layer be (locally) compressed until its width shrinks to $\Delta\sim\lambda_i$. In the ion-inertial region the ions become locally non-magnetic and are accelerated in $-\hat y$ direction by the cross-field electric potential, carrying the pure ion Harris current. Electrons remain magnetised,  transporting the magnetic field with inward velocity $-E/B(z)$ thus giving rise to Hall currents \citep{sonnerup1979} which are restricted solely to the ion-inertial region and close along the magnetic field lines which connect them to the ionosphere \citep{treumann2009}. The $z$-dependence of the Hall currents is 
\begin{displaymath}
J_{\rm H}(z)\approx \frac{eN_0E}{2B_0}\,\frac{\left[1-\Theta\left(|z|-z_i\right)\right]\Theta(|z|-z_e)}{{\rm sinh}(4z/\Delta)}
\end{displaymath}
where $z_{i,e}\equiv\xi_{i,e}\lambda_{i,e}$ and  $1\lesssim\xi_{i,e}\in {\sf R}$ are rational numbers close to but larger than unity. Field line bending in reconnection is not taken into account here. Hall currents vanish in the centre of the current sheet at distances $|z|/\xi_e\leq\lambda_e$, less than the electron inertial length $\lambda_e=c/\omega_e$ where the electrons demagnetise. For the onset of reconnection they are thus of no importance.

The non-magnetic electrons in the central current sheet experience the cross-field potential $\Delta U$, accelerate in $+\hat y$ direction and become the primary carriers of the cross-tail current here. Accelerated to large cross-tail velocities $V_e>v_e$, these electron currents excite the Buneman two-stream instability on growth times shorter than $\tau_{\rm B} < 10^{-3}$ s, a number holding in the magnetotail. The Buneman instability  stabilizes within $0.01< \tau<0.1$ s by heating the trapped electrons along $\pm\hat y$ until $v_{\| e}\sim V_e$. As a consequence the current sheet electrons develop a {\it positive} temperature anisotropy $0<A< 1$ which is sufficiently large to drive the Weibel mode unstable and result in the generation of a stationary magnetic Weibel-field ${\bf B}_{\rm W}=(B_x, 0, B_z)$ in the current sheet with components in the $(x,z)$-plane perpendicular to both, the current flow and anisotropy directions. The fastest growing wavelength is $\lambda_m\sim 2\pi\lambda_e\sqrt{3/A}$. 
\begin{figure*}[t!]
\centerline{\hspace{0cm}{\includegraphics[width=0.5\textwidth,clip=]{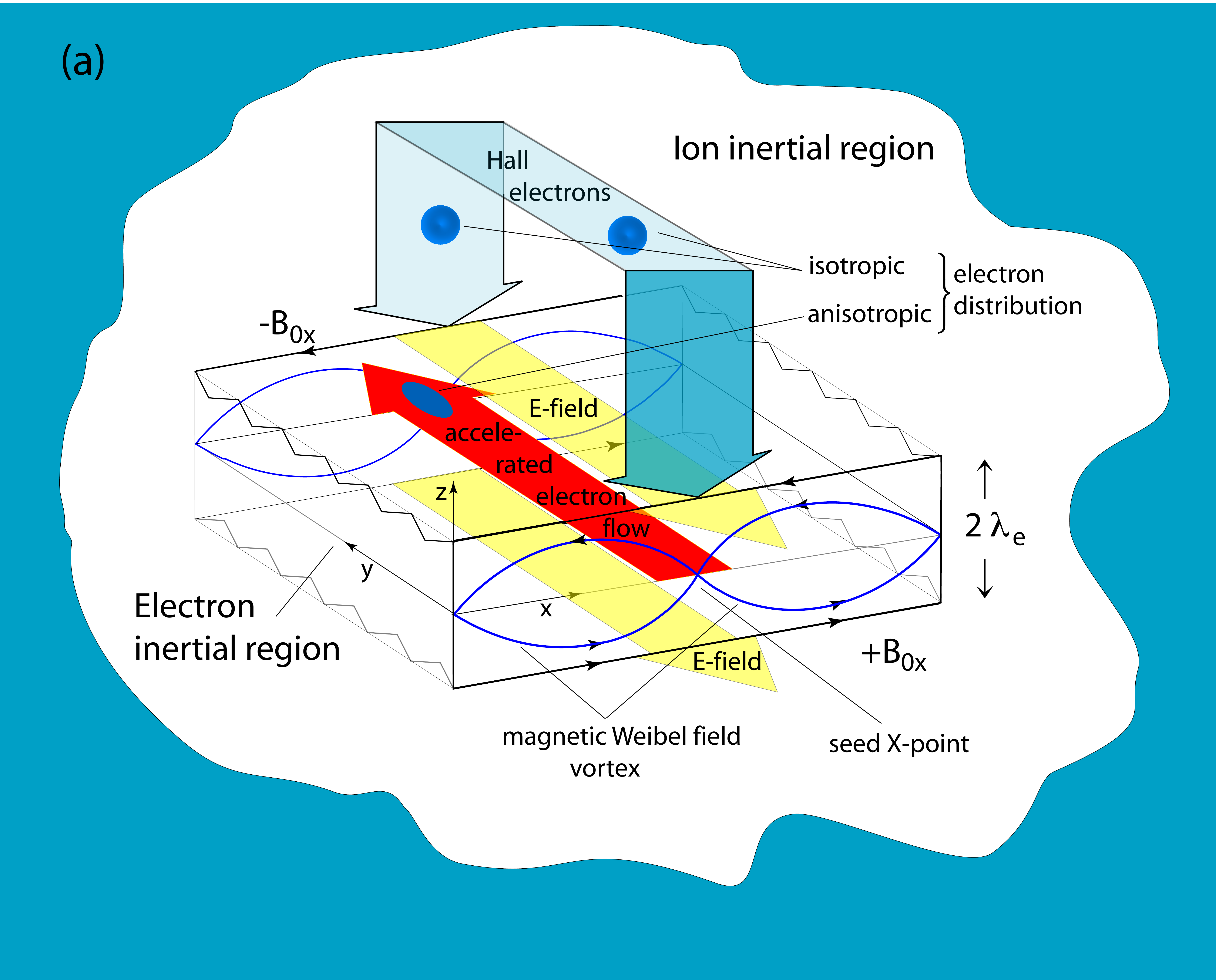} }
\hspace{0cm}{\includegraphics[width=0.5\textwidth,clip=]{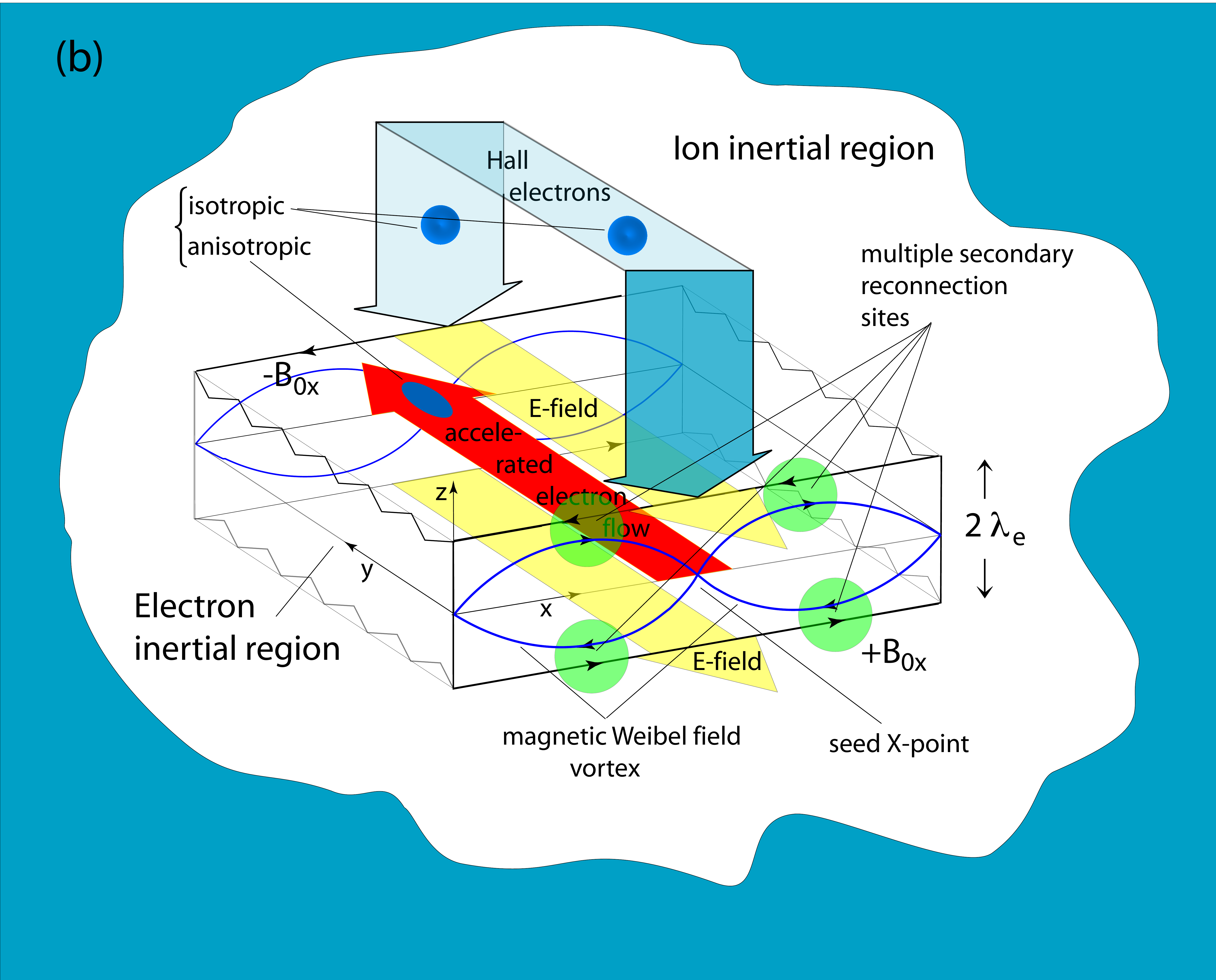} }}
\caption[ ]
{\footnotesize Sketch of the electron-inertial region (width $\Delta z=2\lambda_e$) around the centre of the current layer, embedded into the ion-inertial region. Hall-electrons carrying the Hall-current in the ion-inertial region enter the electron-inertial region (shown on one side of the current layer only) with isotropic temperature distribution, experience the electric field E, accelerate into $+{\hat y}$ direction. After being heated by the two-stream instability they develop a temperature anisotropy and excite magnetic Weibel-vortices (blue) along ${\hat x}$ the vertexes of which serve as seed-{\sf X} points for reconnection. The condition that $B_z=0$ at $z=\pm\lambda_e$ allows for two types of Weibel vortices:  ($a$) a symmetric (magnetically continuous) vortex-mode (same direction as the external fields ${\bf B}_{0x}$), and a  ($b$) anti-symmetric non-continuous vortex-mode (antiparallel to the external field) which yields tail-current bifurcation.The anti-symmetric mode gives rise to multiple reconnection sites  (shown in green colour) and presumably leads to explosive reconnection.}\label{weibel-1}
\vspace{-0.3cm}
\end{figure*}

In the magnetospheric tail current layer we have $\lambda_e\approx 5.4/\sqrt{N_{\rm [cm^{-3}]}}$ km. The maximum growing wavelength of the Weibel magnetic vortices thus becomes shorter than $ \lambda_m<180$ km, the value obtained for a weak anisotropy $A\sim 0.1$. The time for this field to grow to values of the order of 1 nT (or a fraction of it) is of the order of one or few seconds. 

It is usually claimed that the Weibel instability stabilizes when the electron gyroradius in the Weibel magnetic field becomes comparable to the Weibel wavelength. When the electrons are accelerated to $>$ keV energies their gyroradius in a 1~nT magnetic field becomes the order of $\sim 100$ km, roughly the same order as the above estimated maximum wavelength. Thus the short wavelength Weibel field has sufficient time to grow to substantial values until it stabilizes self-consistently by deflecting the current electrons. Prior to this the Weibel field has penetrated the current sheet forming vortices and vertexes which serve as seed-{\sf X} points for reconnection which may then proceed at about along the lines that were discussed long ago in an attempt of formulating a kinetic theory of collisionless  reconnection by  {\it ad hoc} imposing a $B_z$-field on the current layer \citep{galeev1975,sagdeev1979}. This attempt led to the proposal of a scenario for the spontaneous onset of magnetospheric substorms. In the presence of $B_z\neq 0$ the current sheet is in a metastable state that goes spontaneously unstable. The main deficiency of this theory was the lack of any reason for the appearance of $B_z$. Imposing it {\it ad hoc} is the equivalent of igniting reconnection artificially. 

What happens in the long wavelength regime? Here we have $\lambda\gtrsim 300 \lambda_e\approx 1.5\times10^3$ km. The growth time to observable/relevant amplitudes we found to be of the order of  $\tau_W\sim 10$ s, which is not unreasonable for the processes going on there.

The short wavelength modes grow about ten times faster than the long wavelength modes but may not be of substantial importance for reconnection until they cascade inversely down to longer wavelength structures. This could be provided by the coalescence of magnetic islands which is strongest at short wavelengths. The Weibel instability in this case excites an entire spectrum of magnetic field structures in the current layer. In the geomagnetic tail both the short Weibel modes and the long wavelengths  modes  provide seed {\sf X}-points on geophysically reasonable spatial and temporal scales. 

\conclusions[Discussion and Conclusions]
The present approach has so far only implicitly taken into account the narrow width of the non-magnetic central current region which imposes boundary conditions on the evolution of the Weibel mode. Continuity of ${\bf B}$ at $z\sim\pm\lambda_e$ implies a vanishing $B_z$ here. As demonstrated, this imposes limits on wave number and propagation angles of the Weibel mode. In addition the presence of a boundary implies that $B_x$ is either parallel or antiparallel. Clearly the parallel case is preferred as the antiparallel case generates small-scale current bifurcation. On the other hand, this is possible because the evolution of the Weibel mode is completely independent of the presence of the external field.  

One thus distinguishes between two types of Weibel modes depending on their propagation directions $\pm\,{\hat x}$. One of them (Figure \ref{weibel-1}$a$) just causes seed-{\sf X} points, the other (bi- or trifurcated) mode may lead to multiple --  probably explosive --  reconnection (see Figure \ref{weibel-1}$b$). The reason for an explosive character lies in the fact that on the short scale across the boundary $z\sim\pm\lambda_e$ between the inner (Weibel) current region and the external (Hall) region the electrons are unmagnetised such that the Weibel field can freely expand into $z$ until contacting the lobe magnetic field. This causes spontaneous reconnection between the contacting magnetic flux tubes and will force the inner current sheet to decay into a chain of highly dynamical magnetic islands (meso-scale plasmoids), as is immediately realised from Figure \ref{weibel-1}($b$) when imagining the resulting magnetic field structure after multiple spontaneous reconnection has set on. In this case, the inner part of the current layer will become `turbulent' (multiply connected) on meso-scales the order of the Weibel wavelength. The inner current region decays into a `magnetic vortex street' consisting of (electronic) plasmoids and seed-{\sf X} points. 

One may ask what structure of the field and current layer is expected in the direction parallel to the current flow. This question cannot be answered without detailed analysis. However, one may argue that the structure along the current is determined by two facts: the mechanism of electron heating, and the dynamics of the ions. Electron heating occurs in electron holes which have (short) longitudinal extension of $\Delta y\sim 100 \lambda_D$, where $\lambda_D$ is the Debye length. The heating scale is orders of magnitude longer including several to many phase space holes. However, though it is long, it is still microscopic. In the magnetotail current layer this length becomes the order of several $\sim$ 100 km to few 1000 km only. Hence, one suspects that the tail reconnection structure is inherently three-dimensional putting all two-dimensional models in question. 

In addition, the present theory refers to stationary ions. The ions that carry the tail current move into direction $-{\hat y}$. Hence, the Weibel structures and the resulting reconnection sites as a whole move in the direction of the combined speed of the electron holes and current ions. Numerical simulations suggest  \citep[cf., e.g.,][]{newman2002} that this direction is {\it opposite} to the electron flow velocity ${\bf V}_e$, i.e. in the direction of ion flow. As a result one expects that the whole set of magnetotail-reconnection sites will displace slowly -- roughly at translational velocity of the ion-sound speed $c_s\simeq\sqrt{T_e/m_i}\sim200$ km/s -- into $-{\hat y}$-direction, the direction of ion flow, which in the magnetotail is westward. This is in accord with observation of the initial westward displacement of substorm sources. Mapping along the stretched magnetic field lines into the ionosphere decreases this translational westward drift speed by about one order of magnitude. 

In conclusion, it is the thermal-anisotropy driven Weibel instability which provides the magnetic field to penetrate the inner region of the current layer, generates a local normal field component $B_z$ and, by producing short wavelength magnetic vortices and vertexes, it may ignite reconnection on a time scale of tens of seconds to few minutes in the magnetotail. This is in rough agreement with observations, e.g. in the magnetotail, and is sufficiently short for initiating magnetospheric substorms. 

This logically consistent chain of processes provides a satisfactory  mechanism for the spontaneous self-ignited onset of fast magnetic reconnection in thin collisionless current layers. Its numerical verification requires three-dimensional PIC simulations resolving the fully electromagnetic electron dynamics in the current layer.




\begin{acknowledgements}
This research is part of a Visiting Scientist Programme at ISSI, Bern, executed by RT. Hospitality of the ISSI staff is thankfully recognised. Also, the continuous encouragement by Andr\'e Balogh, Director at ISSI, is highly appreciated. 
\end{acknowledgements}











\begin{thebibliography}{ }


\bibitem[Baumjohann et al.(2010)]{baumjohann2010} {Baumjohann, W., Nakamura, R. and Treumann, R. A.}: Magnetic guide field generation in thin collisionless current sheets, {Ann. Geophys.} 28, 789-793, 2010.

\bibitem[Buneman(1958)]{buneman1958} {Buneman, O.}: Instability, turbulence, and conductivity in current-carrying plasma: Phys. Rev. Lett. 1,  8-9, 1958, doi: 10.1103/PhysRevLett.1.8.

\bibitem[Buneman(1959)]{buneman1959} {Buneman, O.}: Dissipation of currents in ionized media: Phys. Rev. 115, 503-517, 1959, doi: 10.1103/PhysRev.115.503.

\bibitem[Dungey(1961)]{dungey1961} 
{Dungey, J.}: Interplanetary magnetic field and the auroral zones, {Phys. Rev. Lett.} {6}, {47-48, doi: 10.1103/PhysRevLett.6.47}, {1961}.

\bibitem[Fried(1959)]{fried1959}
{Fried, B. D.}: Mechanism for instability of transverse plasma waves, {Phys. Fluids} {2}, {337, doi: 10.1063/1.1705933}, {1959}.

\bibitem[Fujimoto et al.(1997)]{fujimoto1997}
Fujimoto, M., Nakamura, M. S., Shinohara, I., Nagai, T., Mukai, T., Saito, Y., Yamamoto, T. and Kokubun, S.: Observations of earthward stre\-aming electrons at the trailing boundary of a plasmoid, Geophys. Res. Lett. 24, 2893-2896, doi: 10.1029/97GL02821, 1997.

\bibitem[Galeev and Zelenyi(1975)]{galeev1975}
Galeev, A. A. and Zelenyi, L. M.: Metastable states of diffuse neutral sheet and the substorm explosive phase,
{J. Exp. Teor. Phys. Lett.} (JETP Lett.) {22}, {170-172}, {1975}.

\bibitem[Nagai et al.(2001)]{nagai2001}   Nagai, T., Shinohara, I., Fujimoto, M., Hoshino, M., Saito, Y., Machida, S. and Mukai, T.:   Geotail observations of the Hall current system: Evidence of magnetic reconnection in the magnetotail,   J. Geophys. Res.   106,  25929-25950, doi: 10.1029/2001JA900038,   2001.

\bibitem[Newman et al.(2001)]
{newman2002}  {Newman, D. L., Goldman, M. V.  and Ergun, R. E.}:  Evidence for correlated double layers, bipolar structures, and very-low-frequency saucer generation in the auroral ionosphere,   {Phys. Plasmas} {9}, {2337-2343, doi: 	10.1063/1.1455004}, {2001}.

\bibitem[{\O}ieroset et al.(2001)]{oieroset2001}   {\O}ieroset, M., Phan, T. D., Fujimoto, M., Lin, R. P. and Lepping, R. P.:   In situ detection of collisionless reconnection in the Earth's magnetotail,   Nature   412,   414-417, doi: 10.1038/35086520,     2001.

\bibitem[Parker(1958)]{parker1958} 
{Parker, E. N.}:  Sweet's mechanism for merging magnetic fields in conducting fluids, {J. Geophys. Res.} {62}, {509-520, doi: 10.1029/JZ062i004p00509}, {1958}.

\bibitem[Sagdeev(1979)]{sagdeev1979}
{Sagdeev, R. Z.}: The 1976 Oppenheimer lectures: Critical problems in plasma astrophysics. I. Turbulence and nonlinear waves, II. Singular layers and reconnection,
{Rev. Mod. Phys.} {51}, {1-20, doi: 	
10.1103/RevModPhys.51.11}, {1979}.

\bibitem[Scudder et al.(2008)]{scudder2008} 
{Scudder, J. D., Holdaway, R. D., Glassberg, R. and Rodriguez, S. L.}:  Electron diffusion region and thermal demagnetization, {J. Geophys. Res.} {113}, {A10208, doi: 10.1029/2008JA013361}, {2008}.

\bibitem[Sitenko(1967)]{sitenko1967}
Sitenko, A. G.: Electromagnetic Fluctuations in Plasma, Academic Press, New York, 1967.

\bibitem[Sonnerup(1979)]{sonnerup1979} Sonnerup, B. U. \"O.: Magnetic Field Reconnection, in:
Solar System Plasma Physics, Vol III, pp. 45-108, eds. L. T. Lanzerotti, C. F. Kennel and  E. N. Parker, North-Holland, New York, 1979.

\bibitem[Sweet(1957)]{sweet1957} 
{Sweet, S.}: The neutral point theory of solar flares, {Proceed. IAU Symp.} {6}, {123-134}, {1958}.

\bibitem[Treumann and Baumjohann(1997)]{treumann1997}
Treumann, R. A. and Baumjohann, W.: Advanced Space Plasma Physics, Imperial College Press, London, 1997.

\bibitem[Treumann et al.(2009)]{treumann2009} 
{Treumann, R., Jaroschek, C. H. and Pottelette, R.}: Auroral evidence for multiple reconnection in the magnetospheric tail plasma sheet, {Europhys. Lett. } 85, {49001, doi: 10.1209/0295 5075/85/49001}, {2009}.

\bibitem[Weibel(1959)]{weibel1959} Weibel, E. S.:  Spontaneously growing transverse waves in a plasma due to an anisotropic velocity distribution,
Phys. Rev. Lett. 2, 83-84, doi: 10.1103/PhysRevLett.2.83, 1959.

\bibitem[Yoon(2007b)]{yoon2007}   Yoon, P. H.:    Spontaneous thermal magnetic field fluctuations,
Phys. Plasmas 14, 064504-064504-4, doi: 10.1063/1.2741388,   2007b.

\bibitem[Yoon and Davidson(1987)]{yoon1987}  Yoon, P. H. and Davidson, R. C.:   Exact analytical model of the classical Weibel instability in a relativistic anisotropic plasma,
Phys. Rev. A 35, 2718-2721, doi: 10.1103/PhysRevA.35.2718, 1987.

\bibitem[Zeiler et al.(2000)]{zeiler2000} {Zeiler, A., Drake, J. F. and Rogers, B. N.}: Magnetic reconnection in toroidal $\eta_i$ mode turbulence,
{Phys. Rev. Lett.} {84}, {99-102, doi: 	
10.1103/PhysRevLett.84.99}, {2000}.

\bibitem[Zurek(2003)]{zurek2003}  Zurek, W. H.:   Coherence, einselection, and the quantum origins of the classical,
Rev. Mod. Phys. 75, 715-775, doi: 10.1103/RevModPhys.75.715.


\end{thebibliography}
\end{document}